\documentclass[12pt]{article}
\usepackage{amsmath}
\usepackage{amssymb}
\renewcommand{\title}[1]{%
    \bigskip%
    \begin{center}%
    \Large\bf #1%
    \end{center}%
    \vskip .2in}

\renewcommand{\author}[1]{%
   {\begin{center}
    #1
    \end{center}}}
\newcommand{\address}[1]{\vspace{-1.7em}\vspace{0pt}
    {\begin{center}
    \it #1
    \end{center}}}

\begin{document}

\title{\bf {Hamiltonian analysis of the Schroedinger field coupled with dynamic non - relativistic gravity }}

\author
{

Pradip Mukherjee  $\,^{\rm a,b}$

Abdus Sattar   $\,^{\rm  c, d}$ }
\address{$^{\rm a}$Department of Physics, Barasat Government College,Barasat, India}
\address{$^{\rm b}$\tt mukhpradip@gmail.com}
 \address{$^{\rm c}$\tt Tapna Chaturbhuj High School\\
Balakhali Tapna, Bishnupur, South 24 pgs, West Bengal-743503, India}
\address{$^{\rm d}$\tt abdussattar9932@gmail.com}
\abstract{ Using the recently mooted Galilean gauge theory we have constructed the model for the Schroedinger field interacting
with gravity which is also dynamical. The dynamics of  gravity is dictated by the Newtonian action in the Newton -
Cartan spacetime. The theory is highly constrained . An elaborate analysis of the constraints of the theory have been 
performed. The symmetries are explicitly verified and the uniqueness of the model has been established. To the best of our 
knowledge both the model and its constraints analysis  are unique in the literature.}

\section {Introduction} 
Non relativistic diffeomorphism invariance  (NRDI) has recently gained considerable interest in the literature \cite{SW,HS,Son,M1, M2,AG} due to its diverse application in condensed matter physics (specifically in the theory of fractional quantum hall effect)(FQHE) \cite{SW}, \cite{M1}, \cite{M2}, holographic models \cite{Janiszewski:2016 auk}, Newtonian Gravity and in many  other 
fields.
  In the Galileo - Newton point of view, gravitation is viewed as an action at a distance. Cartan \cite{Car1} ,  \cite{ Car2}, formulated a geometric approach to non relativistic gravity. This is possible because Newtonian gravity, just as general relativity (GR), satisfies the principle of equivalence  \cite{MTW}. Cartan's geometrical approach to  Newtonian
  gravity manifest as a curvature in a certain space time (later named Newton - Cartan  spacetime ) was developed over a long time \cite{Havas}, \cite{TrautA} , \cite{Kunz}, \cite{Daut},  \cite{EHL},\cite{MALA} . All these developments, however are based 
  on the metric approach with a degenerate metric structure. Recent upsurge of research in this field focuses on the coupling of different non - relativistic dynamic systems with  gravity. For this  a first order vielbein based gravity theory was necessary . Such a theory was mooted only in recent past
  \cite{BMM1}, where non - relativistic diffeomorphism was introduced by gauging the symmetry of a Galilean invariant theory in flat space. Subsequent applications of this methodology and theoretical elaboration of their premises led to a full blown theory of analysing the non - relativistic  diffeomorphism invariant (NRDI) theories , which was christened as Galilean gauge theory (GGT) \cite{BMM1},\cite{ BM1} . The method have explained many a subtle issues in the existing literature  \cite{BMM1},
\cite{ BMM2}, \cite{ BM1}, \cite{ BM3}, \cite{BM4},\cite{ BMN1},\cite{ BMN2},\cite{MS}.  It can safely be claimed that GGT has provided a systematic 
  algorithm for  NRDI , demystifying it, so to say \cite{BGE}.

     In this paper we will consider the  Hamiltonian analysis of a Galilean invariant matter field coupled with Newtonian gravity, applying  GGT . Very recently this was tested \cite{MS} for  Chern - Simons  (C - S) gravity . Since
     the Chern simons action is invariant without reference to any metric \cite{WZ}, the construction of the metric action
     is not that nontrivial, the present work ill be more  interesting , primarily for convenience. However
      We formulate the theory in $(2+1)$ dimensions. Lower dimensional Einstein - Hilbert theories
   have quite distinct properties \cite{carlip}  . So study of the Schroedinger  field  coupled with Newtonian Gravity in $(2+1)$ dimensions will allow us to see whether these characteristics are retained in the non relativistic geometry as well. On the other
   hand the application of NRDI to the strongly correlated electron systems are all in these dimensions. Judging from
  this angle the choice of dimensions is indeed welcome. We should add that the general method described here is applicable in any dimensions.

   It may appear that the most difficult issue is to find action term for Newtonian gravity.To understand how GGT simplifies this analysis  let us compare this with the action in GR, which is the well known Einstein - Hilbert action in Riemann space time. To construct the action we observe  that the field strength tensor (i.e. the commutator of two covariant derivatives is a combination  of the  curvature and torsion. If we put torsion to zero we get GR. Again torsion vanishes when the connection is symmetric. Since GGT provides us the connection in the Newton Cartan space, it can be used to study the geometry.

           The particular gauging of symmetry approach employed here is not new.  Utiyama intrduced a method of gauging the Poincare symmetries of a theory in the Minkowski space \cite{Utiyama}, \cite{Kibble} , \cite{sciama}.  Remarkably, the gauged theory in Minkowski space can be seen as a theory coupled with the Riemann - Cartan spacetime, in the vielbeins formalism .
This theory is called the Poincare gauge theory (PGT). Now we see that there must be a correspondence between PGT and GGT which reflects the correspondence between the Riemann Cartan and Newton Cartan manifolds.
Once we are convinced that such correspondence is possible it is easy to find it \cite{BM5}. Using this correspondence  a particular action,  the Einstein Hilbert action to its limiting form (called the Newton - Cartan action) in the Galilean coordinates can be constructed. Exploiting this correspondence   the complete theory of Schroedinger field coupled with NC back ground  can be obtained.

   Before finishing the introductory section, let us discuss the organisation of the paper. In the next section we have
discussed the non - relativistic Schroedinger field theory, coupled it with Einstein - Hilbert gravity in (2 + 1)
dimensions. In section 3 we provide a faithful Dirac \cite{D}
 analysis of the complete model. The discussion in this part is divided in two subsections. In the first we discussed the
 model building. In The next subsection  constraints analysis is performed. The algebra is given in detail and analysed comprehensively. In the next section our results are provided and discussed. Consistency  of the results is examined in detail. Comparison with the literature reveals an extra ordinary fact, as we will see.                                   s Finally we conclude in section 6.

 \section{ Non relativistic Schrodingeer field coupled wiith non - relativistic gravity}

   The Galilean gauge theory (GGT) enables us to couple a non - relativistic objects like particle, string,   field theory with background gravity \cite{BMM1}, \cite{BMM2}.
The free Schrodinger field theory in Galilean coordinates is given by
\begin{eqnarray}
S = \int d^3x\left[ \frac{i}{2}\left(\psi^*\partial_0\psi - \psi\partial_0\psi^*  \right) -\frac{1}{2m}\partial_k\psi^* \partial_k\psi\right]\label{fs}
\end{eqnarray}
where $\psi$ and $\psi^*$ are the complex Schrodinger fields.By galilean coordinates we mean a set of three number x,y and z calcuated with respetc to some cartesian coordinate system of coordinates. The time is represented by a parameter  $t$, which flows universally. 
The model was introduced in the context of Hall viscosity\cite{SW} in the context of spatial diffeomorphism. When the diffeomorphism is independent of time there is no problem.  Severe problems in gauging are reported when the time dependent transformations creeps in. Galilean transformation 
$\nabla_\mu\psi$ 
where, the greek indices represent the coordinates basis and the latin indices represent the local orthogonal basis which is independently chosen, of the coordinate basis.
Similar considerations apply for $\nabla_0$. So,
\begin{eqnarray}
\nabla_0\psi &=& \Sigma_0{}^\sigma \left(\partial_\sigma + i B_\sigma \right)\psi\nonumber\\
\nabla_a\psi &=&\Sigma a{}^l \left(\partial_l + i B_l \right)\psi\label{kd}
\end{eqnarray}
 $\Sigma$ and $B$ fields, are introduced as compensating (gauge) fields.\cite{BMM2} .Remarkably, the modified theory can be identified with a matter theory coupled with background curvature . $\Sigma$ and $\Lambda$ are comparable with the direct and the inverse vielbein  cosponsoring to their required transformation and the fields $B$ serve as the spin connection of the background geometry. The spin connections of the Newton Cartan spacetime \cite{BMM1,BMM2}, are thus given by,
\begin{equation} 
B_{\mu}=\frac{1}{2}B^{ab}_{\mu}\sigma_{ab}+B^{a0}_{\mu}mx_{a}
\label{bstructure}
\end{equation}
The last equation introduces the independent fields $B^{a0}_{\mu}$ and $B^{ab}_{\mu}$ which, along with 
$\Sigma_\alpha{}^\mu$ , constitute the dynamical variables of the theory. Note that there is an asymmetry in the expression of the covariant derivative,
\begin{eqnarray}\Sigma_a{}^0 & = &0 ; {}{}{}
\Sigma_0{}^k\ne 0 \nonumber\\
B_\mu{}^{0a} & = & 0 {};{}{}B_\mu{}^{a0}\ne 0
\end{eqnarray}
. These are reflection of the fact that time and space are treated in different ways in non - relativistic 
physics. Note that under G. T. the time remains unaltered but the spatial change is also contributed by time Therefore.the
boost generator is represented by the spin connection $B_k{}^{k0}$ . So this is the Galilean boost generator. Note again that $B_k{}^{0k}$is zero.
 At this point we see that the GGT clearly separate the spatial and time sector, Also note that in GGT only the generators of Galilean algebra that actually ensures the invariance of the action (\ref{NR}) are included intrinsically . So we do not need to put any field by hand.

From (\ref{fs}), following the procedure detailed above and correcting for the measure we get the action of Schroedinger field coupled with  background Newtonian gravity.  The Lagrangian density becomes \cite{BMM1}, cite{ BM4}, such that the action is given by,
\begin{eqnarray}
S = \int d^3x \det{\Sigma_\alpha{}^\mu}\left[ \frac{i}{2}\left(\psi^*\nabla_0\psi - \psi\nabla_0\psi^*  \right) -\frac{1}{2m}\nabla_a\psi^* \nabla_a\psi\right]\label{slcompact}
\end{eqnarray}
Note that we still have not included the dynamics of the gravitational field. As we have mentioned earlier this is obtained from GR by the substitutions prescribed in \cite{BM5} for the mapping of PGT to GGT. We will presently go to the issue. For later work we
will write the action in aa way so that the dynamical fields are explicit.
Expanding (\ref{slcompact}), we get
\begin{eqnarray}
\label{ssg}
\mathcal{L}_{1}& =& \frac{M}{\Sigma^{0}_{0}}\Bigl[\frac{i}{2}\Sigma^{0}_{0}\left(\psi^{*}\partial_{0}\psi-\psi\partial_{0}\psi^{*}\right)
\nonumber\\
  & +&\frac{i}{2}\Sigma^{k}_{0}\left(\psi^{*}\partial_{k}\psi-\psi\partial_{k}\psi^{*}\right)
-\Sigma^{\mu}_{a}B^{a0}_{\mu}mx_{a}\psi^{*}\psi\nonumber\\&  -& \frac{1}{2m}\Sigma^{k}_{a}\Sigma^{l}_{a}(\partial_{k}\psi^{*} - i B_K \psi^*)( \partial_{k}\psi + i B_k \psi)
\end{eqnarray}

An important point may be emphasized about the Hamiltonian analysis of (\ref{ssg}). 
In this theory $ \Sigma $ and $B $ are background fields,introduced originally as compensating gauge fields and later identified as the vielbeins and spin connections respectively . From the Hamiltonian point of view these fields act like Lagrange multipliers and not as dynamical fields. They are thus not included in the phase space variables. As a result the symmetries exhibited by the langrangian  does not show up in the Hamiltonian analysis \cite{kluson}.So the dynamic term guiding the motion $\Sigma$ and B must be included in action for the Hamiltonian analysis of the matter coupled with gravity . This is why in the small number of works in the literature gravitional dynamics is excluded \cite{kluson}.
*But the construction of the gravitational actions in the Newton Cartan spacetime is not so difficult. This has been
   made possible by the observations that  under 
   the transformations 
   \begin {equation}
   \Sigma_a{}^0 = 0{}{} {}{};{}{} B_\mu{}^{0a}= 0 
   \label {map}
   \end{equation}

    \cite{BMN1}, the transformation relations of PGT
   become the same as that of GGT \cite{BM5}. We have shown  that any complete dynamics on Riemann (or more generally, Riemann Cartan) spacetime can be reduced to its non - relativistic form
on NC space time using the maps (\ref{map}).. As a verification of this assertion we will first show that the action 
(\ref{ssg}) is invariant under a diffeomorphism $x^\mu \to x^\mu +\xi^\mu$, 
\begin{equation}
\xi^\mu = \epsilon ^\mu + \omega^{\mu}{}_\nu x^\nu - v^a
\label{diff}
\end{equation}



  The 
Lagrangian for the  Einstein Hilbert action is
\begin{equation}
{\mathcal{L}_{EH}}=\epsilon^{\mu\nu\rho}\Lambda^{\alpha}_{\mu}R_{\alpha\nu\rho}
\end{equation}
where
 \begin{eqnarray}
 R_{\alpha\mu\nu}&=&\partial_{\mu}B_{\alpha\nu}-\partial_{\nu}B_{\alpha\mu}+\epsilon_{\alpha\beta\gamma}B_{\beta\mu }B_{\gamma\nu}\nonumber\\
  B_{\alpha\nu}&=-&\frac{1}{2}\epsilon_{\alpha\beta\gamma} B^{\beta\gamma}_{\nu}\nonumber\\
   \end{eqnarray}
Expanding the term we have
\begin{eqnarray}
{\mathcal{L}_{EH}}=\epsilon^{kl}\Lambda^{0}_{0}R_{0kl}+\epsilon^{kl}\Lambda^{a}_{0}R_{akl}-2\epsilon^{kl}\Lambda^{a}_{k}R_{a0l}
\end{eqnarray}
Where
 \begin{eqnarray}
 R_{0kl}&=&\frac{1}{2}\epsilon_{ab}\left(\partial_{l}B^{ab}_{k}-\partial_{k}B^{ab}_{l}+\frac{1}{2}B^{a0}_{k}B^{b0}_{l}\right)\\
  R_{akl}&=&-\frac{1}{2}\epsilon_{ab}\partial_{k}B^{b0}_{l}+\frac{1}{2}\epsilon_{ab}\partial_{l}B^{b0}_{k}-\frac{1}{4}\epsilon_{bc}B^{a0}_{k}B^{bc}_{l}-\frac{1}{2}\epsilon_{bc}B^{ab}_{k}B^{c0}_{l}\\
   R_{a0l}&=&-\frac{1}{2}\epsilon_{ab}\partial_{0}B^{b0}_{l}+\frac{1}{2}\epsilon_{ab}\partial_{l}B^{b0}_{0}-\frac{1}{4}\epsilon_{bc}B^{a0}_{0}B^{bc}_{l}-\frac{1}{4}\epsilon_{bc}B^{ab}_{0}B^{c0}_{l}+\frac{1}{4}\epsilon_{bc}B^{ba}_{0}B^{c0}_{l}
 \end{eqnarray}
  In order to write the appropriate action in the Galilean frame in Newton Cartan spacetime, we have to substitute $\Sigma_a{}^0 = 0$ and $ B_\mu{}^{0a}= 0 $ \cite{BMN1}.

Using the expressions of $R_{0kl}$, $R_{akl}$ and $R_{a0l}$ we can write the E-H piece as,
\begin{eqnarray}
\mathcal{L}_{HE}=\epsilon^{kl}\epsilon_{ab}\bigl[\Lambda^{0}_{0}\partial_{l}B^{ab}_{k}+\frac{1}{4}\Lambda^{0}_{0}B^{a0}_{k}B^{b0}_{l}-\Lambda^{a}_{0}\partial_{k}B^{b0}_{l}+\frac{1}{4}\Lambda^{c}_{0}B^{c0}_{k}B^{ba}_{l}+\frac{1}{2}\Lambda^{c}_{0}B^{cb}_{k}B^{a0}_{l}\nonumber\\+\Lambda^{a}_{k}\partial_{0}B^{b0}_{l}-\Lambda^{a}_{k}\partial_{l}B^{b0}_{0}-\frac{1}{2}\Lambda^{c}_{k}B^{c0}_{0}B^{ba}_{l}-\Lambda^{c}_{k}B^{cb}_{0}B^{a0}_{l}\bigr] \label{NR}
\end{eqnarray}
Which is the desired action for the NR gravity.
 
 Now at this point, one may legitimately  ask about the validity of the term  (\ref{NR}) as  appropriate gravity action in the NC manifold. This amounts to the demonstration that the result of local Galilean transformations at a space time point is a diffieomorphism in the curved space time.Since this result is a very important part of our analysis will show the calculation with great care.
 Let us consider an arbitrary rotation the parameter function of space and time similarly an arbitrary boost the parameter again function of space and time. Considering the special role of time in non relativistic the time translation parameter is taken to be function of time only. This is the local transformation the net result of which the diffeomorphism of $\xi$ which is given by equation (\ref{diff}). The transformation of the fields under the diffeomorphism are obtained from GGT as
\begin{align}
\delta\Lambda^{a}_{0}&=-\xi^{\nu}\partial_{\nu}\Lambda^{a}_{0}-\Lambda^{a}_{\nu}\partial_{0}\xi^{\nu}-v^{a}\Lambda^{0}_{0}\\
\delta\Lambda^{a}_{k}&=-\xi^{\nu}\partial_{\nu}\Lambda^{a}_{k}-\Lambda^{a}_{\nu}\partial_{k}\xi^{\nu}-\omega_{ca}\Lambda^{c}_{k}\\
\delta\Lambda^{0}_{0}&=-\Lambda^{0}_{0}\partial_{0}\xi^{0}-\xi^{\nu}\partial_{\nu}\Lambda^{0}_{0}\\
\delta{B}^{ab}_{\mu}&=-\omega_{ca}B^{cb}_{\mu}-\omega_{cb}B^{ac}_{\mu}-\partial_{\mu}\omega_{ab}-\partial_{\mu}\xi^{\nu}B^{ab}_{\nu}-\xi^{\nu}\partial_{\nu}B^{ab}_{\mu}\\
\delta{B}^{a0}_{\mu}&=-\omega_{ca}B^{c0}_{\mu}+\partial_{\mu}v^{a}-\partial_{\mu}\xi^{\nu}B^{a0}_{\nu}-\xi^{\nu}\partial_{\nu}B^{a0}_{\mu}
\end{align}...
 Using these equations a straightforward but lengthy calculation leads to the following variation of the Lagrangian
\begin{eqnarray}
\delta\mathcal{L} &=& -\xi^{\mu}\partial_{\mu}\mathcal{L_{EH}}-\mathcal{L}_{EH}\partial_{\mu}\xi^\mu-\nonumber \\    
   & &\epsilon^{kl}\Lambda^{0}_{0}v^{a}\left\lbrace \partial_{l}(\epsilon_{ab}B^{b0}_{k})-\epsilon_{db}B^{ad}_{k}B^{b0}_{l}\right\rbrace-\epsilon^{kl}\omega_{db}\Lambda^{a}_{0}   \nonumber\\ & &\left\lbrace \partial_{l}(\epsilon_{ab}B^{d0}_{k})-\epsilon_{cb}B^{ac}_{k}B^{d0}_{l}\right\rbrace 
\end{eqnarray}
for invariance under diffeomorphism must be zero or total derivative.  Now looking at equation apparently the last two term break the invariance but on closure scrutinies we will see that symmetry is unbroaken. The offending term can be written as
\begin{eqnarray}
\delta_o {\mathcal{L}} =
\epsilon^{kl}\Lambda^{0}_{0}v^{a}\left\lbrace \partial_{l}(\epsilon_{ab}B^{b0}_{k})-\epsilon_{db}B^{ad}_{k}B^{b0}_{l}\right\rbrace-\epsilon^{kl}\omega_{db}\Lambda^{a}_{0}  \left\lbrace \partial_{l}(\epsilon_{ab}B^{d0}_{k})-\epsilon_{cb}B^{ac}_{k}B^{d0}_{l}\right\rbrace 
\label{offending}
\end{eqnarray}
We start with the coefficient of $v^a$ as
\begin{equation}
\epsilon^{kl} \Lambda ^0{}_0 \left(\epsilon_{ab} \partial_l B_k{}^{b 0} + B_l{}^{a d}\epsilon_{d b}B_k{}^{b 0}\right)
\label{term}
\end{equation}

Now we are working in the Galilean coordinates where the constant time slices are flat. So the parallel transport of a
geometric quantity along a space direction does not change it. Since $B_k{}^{b 0} $ contain both local and coordinate indices ,
\begin{eqnarray}
 D_ l\left( B + \Gamma  \right)\epsilon_{a b}B_k{}^{b 0} = 0
 \end{eqnarray}

 Expanding the last equation, 
\begin{eqnarray}
 \partial_l (\epsilon_{ab} B_k{}^{b o} + B_l {}^{b d} 
                                  \epsilon_{db} B_k{}^{b0} - \Gamma^{m}{}_{kl} \epsilon_{db} B_m{}^{b0} =0
                                  \label{inv}
\end{eqnarray}                                                            
 Interchanging $k$ and $l$ 
\begin{eqnarray}
 \partial_l (\epsilon_{ab} B_l{}^{b o}) + B_k {}^{b d} 
                                  \epsilon_{db} B_l{}^{b0} - \Gamma^{m}{}_{lk} \epsilon_{db} B_m{}^{b0} =0
                                  \label{inv1}
\end{eqnarray}    
 Since the spacetime is torsionless   $ \Gamma^{m}{}_{kl} = \Gamma^{m}{}_{lk} $ . Subtracting (\ref{inv1})  from
 (\ref{inv}) , we obtain

\begin{eqnarray}
 \left( \partial_l (\epsilon_{ab} B_k{}^{b 0} - \partial_k (\epsilon_{ab} B_k{}^{b 0}\right)
                 + \left(B_l {}^{b d} 
                                  \epsilon_{db} B_k{}^{b0} - B_l {}^{b d} 
                                  \epsilon_{db} B_k{}^{b0}\right) =0
                                   \end{eqnarray} 

The last equation can be written as,

\begin{eqnarray}
 \epsilon^{kl}\left( \left( \partial_l (\epsilon_{ab} B_k{}^{b 0}\right) - \partial_k (\epsilon_{ab} B_l{}^{b 0} \right) =0
 \end{eqnarray}                                                                       = 0

\begin{eqnarray}
\delta\mathcal{L}=-\xi^{\mu}\partial_{\mu}\mathcal{L_{EH}}-\mathcal{L}_{EH}\partial_{\mu}\xi^\mu
\end{eqnarray}
we would like to verify whether the arbitrary local rotation and boost plus coordinate transformation keeps the action (equation) invarianct.
 Take the first term
 \begin{equation}
 =\epsilon^{kl}\epsilon_{ab}\Lambda^{0}_{0}\partial_{l}B^{ab}_{k}
 \end{equation}
 Replace it by 
 \begin{equation}
 =\epsilon^{kl}\epsilon_{ab} (\Lambda^{0}_{0} + \delta\Lambda^{0}_{0})
 \left[\partial_{l}(B^{ab}_{k} + \delta B^{ab}_{k})\right]  
 \end{equation}
 
 we can now conclude that the dynamically complete  Lagrangian density is given by
\begin{equation}
\mathcal{L} = \mathcal{L}_{1}+\mathcal{L}_{EH}  \label{first} \\
\end{equation}. 
and, 
 
Consistency in the Hamiltonian analysis is essential for a feasible model. We will see that the 
model (\ref{first}) for the Schroedinger field coupled with non relativistic space is consistent from this point of view. This is remarkable because a host of models have been proposed for this problem, many of which have some differences with (\ref{first}). Also it may be pointed out that Hamiltonian treatment of these theories are not much available.  

 In the following section we will discuss the Dirac approach to the constraint analysis of the problem.

\section{Canonical Analysis - the constraints of the theory}

 Hamiltonian analysis of the theories of gravitation  (like General relativity) is very important as well as
 really interesting  
as these theories are already covariant. Diffeomorphism invariance of the theory  endows it with a plethora of symmetries which are organically related with the phase space constraints. In case of non - relativistic diffeomorphism invariance (NRDI) examples of this kind of 
 Hamiltonian analysis are rare. So we will provide details of the calculation wherever we feel it suitable, running the
 the risk of quoting quite ugly mathematical manipulations . We may hope that this cavalier task will ultimately be   amply rewarded 
 
 \subsection{Constraints of the model}

To proceed with the canonical analysis of (\ref{NR}) we define the momenta $\pi$, $\pi^{*}$,$\pi^{0}_{\mu}$, $\pi^{a}_{k}$, $\pi^{\mu}_{ab}$, $\pi^{l}_{b0}$, $\pi^{0}_{a0}$,  conjugate to the fields $\psi$, $\psi^{*}$,$\Sigma^{\mu}_{0}$, $\Sigma^{k}_{a}$, $B^{ab}_{\mu}$, $B^{b0}_{l}$, $B^{a0}_{0}$ respectively. Then
\begin{eqnarray}
\pi &=&\frac{\partial\mathcal{L}}{\partial\dot{\psi}} =\frac{Mi}{2}\psi^{*} \hspace{.2cm}
\pi^{*}= \frac{\partial\mathcal{L}}{\partial\dot{\psi^{*}}} =-\frac{Mi}{2}\psi \nonumber \\
\pi^{0}_{\mu}&=&\frac{\partial\mathcal{L}}{\partial\dot{\Sigma^{\mu}_{0}}}=0 
\pi^{a}_{k} =\frac{\partial\mathcal{L}}{\partial\dot{\Sigma^{k}_{a}}}=0  \hspace{.2cm};\nonumber \\
\pi^{\mu}_{ab}&=&\frac{\partial\mathcal{L}}{\partial\dot{B^{ab}_{\mu}}}=0 
\pi^{l}_{b0} =  \frac{\partial\mathcal{L}}{\partial{\dot{B^{b0}}_{l}}}=
\epsilon^{kl}\epsilon_{ab}\Lambda^{a}_{k} \nonumber \\
\pi^{0}_{a0} &=& \frac{\partial{\mathcal{L}}}{\partial{\dot{B^{a0}}_{0}}}= 0
\label{m}
\end{eqnarray}
The Poisson brackets (PB) between the canonical pairs are usual:
\begin{eqnarray}
{\{\psi(x),\pi(y)\}}=\delta^{2}(x-y)\notag\\
{\{\psi^{*}(x),\pi^{*}(y)\}}=\delta^{2}(x-y)\notag\\
{\{\Sigma^{\mu}_{0}(x),\pi^{0}_{\nu}(y)\}}=\delta^{\mu}_{\nu}\delta^{2}(x-y)\notag\\
{\{\Sigma^{l}_{b}(x),\pi^{a}_{k}(y)\}}=\delta^{a}_{b}\delta^{l}_{k}\delta^{2}(x-y)\notag\\
{\{B^{ab}_{\mu}(x),\pi^{\nu}_{cd}(y)\}}=\delta^{\nu}_{\mu}(\delta^{a}_{c}\delta^{b}_{d}-\delta^{b}_{c}\delta^{a}_{d})\delta^{2}(x-y\notag\\
{\{B^{a0}_{k}(x),\pi^{l}_{b0}(y)\}}=\delta^{l}_{k}\delta^{a}_{b}\delta^{2}(x-y)\notag\\
{\{B^{b0}_{0}(x),\pi^{0}_{a0}(y)\}}=\delta^{b}_{a}\delta^{2}(x-y)
 \label{cpb}
\end{eqnarray}
All these definitions lead to  primary constraints. 
From definition (\ref{m}) the following primary constraints emerge,
\begin{align}
\Omega_{1}=\pi-\frac{Mi}{2}\psi^{*}\thickapprox 0\hspace{.2cm};\hspace{.2cm}
\Omega_{2}&=\pi^{*}+\frac{Mi}{2}\psi \thickapprox 0\notag\\
\Omega^{0}_{\mu}=\pi^{0}_{\mu}\thickapprox 0\hspace{.2cm};\hspace{.2cm}
\Omega^{a}_{k}&=\pi^{a}_{k}\thickapprox 0\notag\\
\Omega^{\mu}_{ab}=\pi^{\mu}_{ab}\thickapprox 0\hspace{.2cm};\hspace{.2cm}
\Omega^{l}_{b0}&=\pi^{l}_{b0}-\epsilon^{kl}\Lambda^{a}_{k}\epsilon_{ab} \thickapprox 0\hspace{.2cm}\notag\\\Omega^{0}_{a0}&=\pi^{0}_{a0}\thickapprox 0 
\label{pc1}
\end{align}
As is well known, conserving the primary constraints (\ref{pc1}) we may get secondary constraints.
We have to construct the total Hamiltonian, which is the canonical Hamiltonian improved by the linear combinations of the primary constraints.
The canonical Hamiltonian density of the theory is given by
\begin{eqnarray}
\mathcal{H}_{can}=\pi\dot{\psi}+\pi^{*}\dot{\psi^{*}}+\pi^{0}_{\mu}\dot{\Sigma^{\mu}_{0}}+\pi^{a}_{k}\dot{\Sigma^{k}_{a}}+\pi^{\mu}_{ab}\dot{B^{ab}_{\mu}}+\pi^{l}_{b0}\dot{B^{b0}_{l}}+\pi^{0}_{a0}\dot{B^{a0}_{0}}-\mathcal{L}
\label{canh}
\end{eqnarray}
Explicitly,
\begin{multline}
\mathcal{H}_{can}=-\frac{M}{\Sigma^{0}_{0}}\Bigl[\frac{i}{2}\Sigma^{k}_{0}\left(\psi^{*}\partial_{k}\psi-\psi\partial_{k}\psi^{*}\right)-\Sigma^{\mu}_{0}B^{a0}_{\mu}mx_{a}\psi^{*}\psi\\
-\frac{1}{2m}\Sigma^{k}_{a}\Sigma^{l}_{a}\left(\partial_{k}\psi^{*}\partial_{l}\psi+iB^{b0}_{l}mx_{b}\psi\partial_{k}\psi^{*}-iB^{b0}_{k}mx_{b}\psi^{*}\partial_{l}\psi+B^{c0}_{k}B^{b0}_{l}m^{2}x_{c}x_{b}\psi^{*}\psi\right)\Bigr]\\
-\epsilon^{kl}\epsilon_{ab}\bigl[\Lambda^{0}_{0}\partial_{l}B^{ab}_{k}+\frac{1}{4}\Lambda^{0}_{0}B^{a0}_{k}B^{b0}_{l}-\Lambda^{a}_{0}\partial_{k}B^{b0}_{l}+\frac{1}{4}\Lambda^{c}_{0}B^{c0}_{k}B^{ba}_{l}+\frac{1}{2}\Lambda^{c}_{0}B^{cb}_{k}B^{a0}_{l}\nonumber\\-\Lambda^{a}_{k}\partial_{l}B^{b0}_{0}-\frac{1}{2}\Lambda^{c}_{k}B^{c0}_{0}B^{ba}_{l}-\Lambda^{c}_{k}B^{cb}_{0}B^{a0}_{l}\bigr]
\end{multline}

The total Hamiltonian is

\begin{multline}
H_{T}=\int{d^{2}x}\bigl[  \mathcal{H}_{can}+\lambda_{1}\Omega_{1}+\lambda_{2}\Omega_{2}+\lambda^{\mu}_{0}\Omega^{0}_{\mu}+\lambda^{k}_{a}\Omega^{a}_{k}+\frac{1}{2}\lambda^{ab}_{\mu}\Omega^{\mu}_{ab}
+\lambda^{b0}_{l}\Omega^{l}_{b0}+\lambda^{a0}_{0}\Omega^{0}_{a0}\bigr] 
\end{multline}

Here $\lambda_{1}$, $\lambda_{2}$, $\lambda^{\mu}_{0}$, $\lambda^{k}_{a}$, $\lambda^{ab}_{l}$, $\lambda^{b0}_{l}$, $\lambda^{a0}_{0}$ are Lagrange multipliers enforcing the constraints. In this theory, the non-vanishing fundamental Poisson brackets are given by
\begin{align*}
\label{pc1}
{\{\Omega_{1}(x),\Omega_{2}(y)\}}&=-iM\delta^{2}\left(x-y\right)\\
{\{\Omega_{1}(x),\Omega^{a}_{k}(y)\}}&=\frac{i\psi^{*}}{2}M\Lambda^{a}_{k}\delta^{2}\left(x-y\right)\\
{\{\Omega_{2}(x),\Omega^{a}_{k}(y)\}}&=-\frac{i\psi}{2}M\Lambda^{a}_{k}\delta^{2}\left(x-y\right)\\
{\{\Omega^{a}_{k}(x),\Omega^{l}_{b0}(y)\}}&=-\epsilon^{jl}\epsilon_{cb}\Lambda^{c}_{k}\Lambda^{a}_{j}\delta^{2}\left(x-y\right)
\end{align*}
where we have used (\ref{cpb}). The primary constraints are denoted by the generic symbol $\Omega$, The index structure is sufficient to identify the particular one of the set , if necessary. . Apparently, all the constraints
have nonzero PBs between each other, However, it may so happen that by a  combination of the
constraints, a subset of them can be formed , which have vanishing PBs with all the elements of the set of constraints. For the time being let us carry on with 
the stationary of the primary constraints $\Omega^{0}_{d0}$ i.e; $\dot{\Omega}^{0}_{d0}={\{\Omega^{0}_{d0}(x),H_{T}\}}.
 \thickapprox 0$ which yields the 
\begin{equation}
\label{s15}
-Mmx_{d}\psi^{*}\psi+\epsilon^{kl}\epsilon_{ad}\partial_{l}\left(\Lambda^{a}_{k}\right)+\frac{1}{2}\epsilon^{kl}\epsilon_{ab}\Lambda^{d}_{k}B^{ab}_{l} \thickapprox 0 
\end{equation}
Note that the condition (\ref{s15})should hold for all values of $x^d $, terms containing $x^d$ and the rest are separately zero. Two new secondary constraints are,
thus obtained,
\begin{eqnarray}
\label{s}
\Phi= \psi^{*}\psi\thickapprox 0 
\end{eqnarray}
and
\begin{equation}
\label{s11}
\Phi_{d}=\epsilon^{kl}\epsilon_{ad}\partial_{l}\left(\Lambda^{a}_{k}\right)+\frac{1}{2}\epsilon^{kl}\epsilon_{ab}\Lambda^{d}_{k}B^{ab}_{l} \thickapprox 0 
\end{equation}

The stationary of the primary constraint $\Omega^{0}_{ef}$ i.e; $\dot{\Omega}^{0}_{ef}={\{\Omega^{0}_{ef}(x),H_{T}\}} \thickapprox 0$ gives the secondary constraints as

\begin{equation}
\Gamma_{ef} =\epsilon^{kl}\epsilon_{ab}\Lambda^{c}_{k}B^{b0}_{l}\left(\delta^{c}_{e}\delta^{a}_{f}-\delta^{a}_{e}\delta^{c}_{f}\right)  \thickapprox 0
\end{equation}
Conserving $\Omega^{j}_{ef}$ in time, a secondary constraint emerges

\begin{equation}
\Gamma^{j}_{ef}=\epsilon^{kl}\epsilon_{ab}\left[\partial_{l}(\Lambda^{0}_{0})\delta^{j}_{k}\left(\delta^{a}_{e}\delta^{b}_{f}-\delta^{b}_{e}\delta^{a}_{f}\right)-\frac{1}{2}\Lambda^{c}_{k}B^{c0}_{0}\delta^{j}_{l}\left(\delta^{a}_{e}\delta^{b}_{f}-\delta^{b}_{e}\delta^{a}_{f}\right)\right]  \thickapprox0
\end{equation}
Conserving $\Omega^{0}_{0}$ in time, two secondary constraints emerge
\begin{equation}
S_{k}=\left(\psi\partial_{k}\psi^{*}-\psi^{*}\partial_{k}\psi^{*}\right)\thickapprox 0
\end{equation}
\begin{equation}
S=\frac{M}{2m}\Sigma^{k}_{a}\Sigma^{l}_{a}\partial_{k}\psi^{*}\partial_{l}\psi-\epsilon^{kl}\epsilon_{ab}\left( \partial_{l}B^{ab}_{k}+\frac{1}{4}B^{a0}_{k}B^{b0}_{l}\right)\thickapprox 0   
\end{equation}
Finally, conservation of $\Omega^{0}_{j}$ leads to secondary constraint 
\begin{equation}
\Gamma_{j}=\epsilon ^{kl}\epsilon_{ab}\left(\Lambda^{a}_{j} \partial_{l}B^{b0}_{k}-\frac{1}{4}\Lambda^{c}_{j}B^{c0}_{k}B^{ab}_{l}-\frac{1}{2}\Lambda^{c}_{j}B^{ca}_{k}B^{b0}_{l}\right)\thickapprox 0
\end{equation}
Conserving the rest of the primary constraints $\Omega_1$, $\Omega_2$, $\Omega^a_k$, $\Omega^l_{b0}$ and the
new secondary constraints $\Phi$, $\Phi_{d}$, $\Phi_{ef}$, $\Phi^{j}_{ef}$, $\Phi_{j}$, $S_{K}$, $S$    no new constraints generate ; only some of the multipliers are fixed. The number of the the constraints in the system is now closed. 

The secondary constraints obtained in the above are  listed below for ready reference,:
\begin{align}
\Phi&= \psi^{*}\psi\thickapprox 0 \notag \\
\Phi_{d}&=\epsilon^{kl}\epsilon_{ad}\partial_{l}\left(\Lambda^{a}_{k}\right)+\frac{1}{2}\epsilon^{kl}\epsilon_{ab}\Lambda^{d}_{k}B^{ab}_{l} \thickapprox 0\notag \\
\Gamma_{ef}&=\epsilon^{kl}\epsilon_{ab}\Lambda^{c}_{k}B^{b0}_{l}\left(\delta^{c}_{e}\delta^{a}_{f}-\delta^{a}_{e}\delta^{c}_{f}\right)  \thickapprox 0\notag \\
\Gamma^{j}_{ef}&=\epsilon^{kl}\epsilon_{ab}\left[\partial_{l}(\Lambda^{0}_{0})\delta^{j}_{k}\left(\delta^{a}_{e}\delta^{b}_{f}-\delta^{b}_{e}\delta^{a}_{f}\right)-\frac{1}{2}\Lambda^{c}_{k}B^{c0}_{0}\delta^{j}_{l}\left(\delta^{a}_{e}\delta^{b}_{f}-\delta^{b}_{e}\delta^{a}_{f}\right)\right]  \thickapprox0\notag \\
S_{k}&=\left(\psi\partial_{k}\psi^{*}-\psi^{*}\partial_{k}\psi^{*}\right)\thickapprox 0 \notag \\
S&=\frac{M}{2m}\Sigma^{k}_{a}\Sigma^{l}_{a}\partial_{k}\psi^{*}\partial_{l}\psi-\epsilon^{kl}\epsilon_{ab}\left( \partial_{l}B^{ab}_{k}+\frac{1}{4}B^{a0}_{k}B^{b0}_{l}\right)\thickapprox 0\notag \\
\Gamma_{j}&=\epsilon ^{kl}\epsilon_{ab}\left(\Lambda^{a}_{j} \partial_{l}B^{b0}_{k}-\frac{1}{4}\Lambda^{c}_{j}B^{c0}_{k}B^{ab}_{l}-\frac{1}{2}\Lambda^{c}_{j}B^{ca}_{k}B^{b0}_{l}\right)\thickapprox 0 
\end{align}
Set of constraints is $\Phi$,$\Phi_{d}$,$\Gamma_{ef}$,$\Gamma^{j}_{ef}$,$S_{k}$,$S$. Our set of constraints can be classified into first and second sector.This will be done in following subsection. 

\subsection{Classification  of the  constraints   and degrees of freedom count} In the Dirac method the constraints are divided in first and second class according to whether they have all mutual Poisson brackets vanishing or not. Using the fundamental Poisson brackets (\ref{cpb}) we can straightforwardly work out these brackets. 
The non-vanishing Poisson brackets are given by-
\begin{align}
\label{split}
{\{{\Omega}_{1}(x),\Omega_2(y)\}}&=-iM\delta^{2}(x-y)\\
{\{\Omega_{1}(y)},\Omega^{a}_{k}(x)\}&=\frac{i\psi^{*}}{2}M\Lambda^{a}_{k}\delta^{2}(x-y)\\
{\{\Omega_{1}(x),\Phi(y)}\}&=-\psi^{*}\delta^{2}(x-y)\\
{\{\Omega_{1}(x),S_{k}(y)}\}&=\left[\psi^{*}(y)\partial_{k}^{y}(\delta^{2}(x-y))-\partial_{k}^{y}\psi^*\delta^{2}(x-y)\right]\\
{\{\Omega_{1}(x),S(y)}\}&=-\frac{M}{2m}\Sigma^{k}_{a}\Sigma^{l}_{a}\partial_{k}^{y}\psi^*\partial_{l}^{y}\left(\delta^2(x-y)\right)\bigr]\\ 
{\{\Omega_{2}(y)},\Omega^{a}_{k}(x)\}&=-\frac{i\psi}{2}M\Lambda^{a}_{k}\delta^{2}(x-y)\\
{\{\Omega_{2}(x),\Phi(y)}\}&=-\psi\delta^{2}(x-y)\\
{\{\Omega_{2}(x),S_{k}(y)}\}&=-\left[\psi(y)\partial_{k}^{y}(\delta^{2}(x-y))+\partial_{k}^{y}\psi\delta^{2}(x-y)\right]\\
{\{\Omega_{2}(x),S(y)}\}&=-\frac{M}{2m}\Sigma^{k}_{a}\Sigma^{l}_{a}\partial_{l}^{y}\psi\partial_{k}^{y}\left(\delta^2(x-y)\right)\bigr]
\end{align}
\begin{align} 
\label{split1}\\
{\{\Omega^{0}_{0}(x)},\Gamma^{j}_{ef}(y)\}&=\epsilon^{lj}\epsilon_{ab}\left(\delta^{d}_{e}\delta^{b}_{f}-\delta^{b}_{e}\delta^{d}_{f}\right)\partial_{l}^{y}\left((\Lambda^{0}_{0})^2\delta^{2}(x-y)\right)\notag\\
{\{\Omega^{a}_{k}(x)},\Omega^{l}_{b0}(y)\}&=-\epsilon^{pl}\epsilon_{db}\Lambda^{d}_{k}\Lambda^{a}_{p}\delta^2(x-y)\\
 {\{\Omega^{a}_{k}(x)},\Gamma_{ef}(y)\}&=\epsilon^{jl}\epsilon_{db}\Lambda^{c}_{k}\Lambda^{a}_{j}B^{b0}_{l}\left( \delta^{c}_{e}\delta^{d}_{f}-\delta^{d}_{e}\delta^{c}_{f}\right) \delta^2(x-y)\\
 {\{\Omega^{a}_{k}(x)},\Gamma^{j}_{ef}(y)\}&=-\frac{1}{2}\epsilon^{pj}\epsilon_{db}\Lambda^{c}_{k}\Lambda^{a}_{p}B^{c0}_{0}\left(\delta^{d}_{e}\delta^{b}_{f}-\delta^{b}_{e}\delta^{d}_{f}\right)\delta^2(x-y)\\
{\{\Omega^{a}_{k}(x)},\Gamma_{j}(y)\}&=\epsilon^{pl}\epsilon_{db}\left[ \Lambda^{d}_{k}\Lambda^{a}_{j}\partial_{l}B^{b0}_{p}-\frac{1}{4}\Lambda^{c}_{k}\Lambda^{a}_{j}B^{c0}_{p}B^{db}_{l}-\frac{1}{2}\Lambda^{c}_{k}\Lambda^{a}_{j}B^{cd}_{p}B^{b0}_{l}\right] \delta^2(x-y)\\
{\{\Omega^{a}_{k}(x)},S(y)\}&=\frac{M}{2m}\partial_{p}\psi^*
\partial_{l}\psi\left[\Sigma^{p}_{c}\Sigma^{l}_{c}\Lambda^{a}_{k}-\Sigma^{l}_{a}\delta^{p}_{k}-\Sigma^{p}_{a}\delta^{l}_{k}\right] \delta^2(x-y)\notag\\
\label{split2}\\
{\{\Omega^{0}_{a0}(x)},\Gamma^{j}_{ef}(y)\}&=\frac{1}{2}\epsilon^{pj}\epsilon_{db}\Lambda^{a}_{p}\left(\delta^{d}_{e}\delta^{b}_{f}-\delta^{b}_{e}\delta^{d}_{f}\right)\delta^2(x-y)\\
{\{\Omega^{k}_split{ab}(x)},\Phi_{d}(y)\}&=-\frac{1}{2}\epsilon^{pk}\epsilon_{cf}\Lambda^{d}_{p}\left(\delta^{c}_{a}\delta^{f}_{b}-\delta^{f}_{a}\delta^{c}_{b}\right)\delta^2(x-y)\\
 {\{\Omega^{k}_{ab}(x)},\Gamma_{j}(y)\}&=\frac{1}{4}\epsilon^{pk}\epsilon_{df}\Lambda^{c}_{j}B^{c0}_{p}\left(\delta^{d}_{a}\delta^{f}_{b}-\delta^{f}_{a}\delta^{d}_{b}\right)\delta^2(x-y)+\frac{1}{2}\epsilon^{kl}\epsilon_{df}\Lambda^{c}_{j}B^{f0}_{l}\left(\delta^{c}_{a}\delta^{d}_{b}-\delta^{d}_{a}\delta^{c}_{b}\right)\delta^2(x-y)\\
 {\{\Omega^{k}_{ab}(x)},S(y)\}&=\epsilon^{kl}\epsilon_{cd}\left(\delta^{c}_{a}\delta^{d}_{b}-\delta^{d}_{a}\delta^{c}_{b}\right)\partial^{y}_{l}(\delta^2(x-y))\\
{\{\Omega^{l}_{b0}(x)},\Gamma_{ef}(y)\}&=-\epsilon^{pl}\epsilon_{ab}\Lambda^{c}_{k}\left(\delta^{c}_{e}\delta^{a}_{f}-\delta^{a}_{e}\delta^{c}_{f}\right)\delta^2(x-y)\\
{\{\Omega^{l}_{b0}(x)},\Gamma_{j}(y)\}&=-\epsilon^{kp}\epsilon_{ad}\left[ \Lambda^{a}_{j}\delta^{l}_{k}\delta^{d}_{b}\partial^{y}_{l}(\delta^2(x-y))-\frac{1}{4}\Lambda^{b}_{j}\Lambda^{l}_{k}B^{ad}_{p}\delta^2(x-y)-\frac{1}{2}\Lambda^{c}_{j}\delta^{l}_{p}\delta^{d}_{b}B^{ca}_{k}\delta^2(x-y)\right]\\
{\{\Omega^{l}_{b0}(x)},S(y)\}&=\frac{1}{2}\epsilon^{kl}\epsilon_{ab}B^{a0}_{k}\delta^2(x-y)\\
\end{align}
From above PBs bracket we find that $\Omega^{0}_{ab}$ and $\Omega^{0}_{k}$ have vanishing PBs with other constraints.So this constraint is first class.
The complete classification of constraints is summarized in Table \ref{tab:Classification of Constraints} below.
\begin{table}[ht]
\caption{Classification of Constraints}
\label{tab:Classification of Constraints}
\centering
\begin{tabular}{c c c}
\hline\hline\\
\ & First Class & Second Class \\
 \hline\\
 Primary &  ${\Omega^{0}_{ab}}$, $\Omega^{0}_{k}$ & ${\Omega_{1}}$,${\Omega_{2}}$, ${\Omega^{0}_{0}}$, $\Omega^{a}_{k}$, $\Omega^{0}_{a0}$, $\Omega^{l}_{b0}$, $\Omega^{k}_{ab}$ \\[1em]
\hline\\
 Secondary &  &  $\Phi$, $\Phi_{d}$, $\Gamma_{ef}$, $\Gamma^{j}_{ef}$, $\Gamma_{j}$, $S$, $S_{k}$, \\[1em]
 \hline\hline
\end{tabular}
\end{table}

  Every iotam of the Hamiltonian anlysis given above are new. So we require the physical significance of these  results.
  This topic will be taken up in the following section.
  
    \section { Discussion of the results}

    We begin with the consistency of  the canonical  analysis  given above. For convenience the discussion will be 
    organised  in several points.

\begin{enumerate}
\item

 The number of fields is 18. That gives 36 fields in the phase space as each field is accompanied with its canonically conjugate momentum. Not all of them are independent due to the phase space constraints. A second class constraint reduces the number of degrees of freedom by one while a first class constraint reduce the number by 2.  The number of first class constraints here is 3 while the number of second class constraints is 26. The number of independent degrees of freedom in the phase space, $N = 36 - 2 \times 3 - 26 =4  $. So, the no. of degrees of freedom in configuration space is 2. This is the expected result as physically, they correspond to $\psi$ and $\psi^*$. Note that the Einstein Hilbert dynamics in (2 + 1) dimensions does not contribute any propagating degree of freedom. \cite {witten}. 
 
 \item 
 
 The number of independent primary first class constraints is three. According to Dirac conjecture , it is the number of independent 'gauge' degrees of freedom. Here arbitrary functions in the solutions of the equations of motion will then be three in number. Physically, these are the consequence of three local symmetry operations, one rotation and two boosts.
 The close connection between the theoretical and physical results are in unison in GGT, which is the prime achievement  
 of our theory.
\end{enumerate}

Now, in relativistic realm when also the Einstein Hilbert action is quite conspicuous. Thus, only in these dimensions, the
theory js equivalent to a pure gauge theory\cite{witten}. Again it has no independent dynamics. Our results show that this also true for the non - relativistic theories.

Now we will impose another check on our model. The theory in its usual formulation \cite{jensen} does not contain
$\Sigma_0{}^a$  and instead has an additional $U(1)$ gauge fild. We would verify the result of a similar model (but with no gauge field) whether it is consistent or not. 

  \section{Consistency of the Galilean gauge theory}

In the above we have given the Hamiltonian analysis of the non relativistic  Schroedinger field coupled   with back ground Newtonian gravity.  It may be observed that this is substantially facilitated by the Galilean gauge theory. The same Schroedinger field has been discussed by many authors. Chronologically, \cite{SW} is the earliest
We have already discussed at few places in this paper that the motive of our work is to check the consistency of the model (\ref{NR}) and to posit it in relation to the corresponding actions obtained from other approaches. To our knowledge the latter  are of the  same form as that of \cite{SW}. This form differs from our model in essence by the absence of the term $ \Sigma_0{}^k$ which is taken to b zero. It will then be crucial to check what happens when in our model we substitute $\Sigma_0{}^k$ = 0 i.e. is it still has the same
physically consistent Hamiltonian structure.

 We therefore consider the truncated model
 \begin{equation}
\mathcal{L} = \mathcal{L}_{1}+\mathcal{L}_{EH}
\end{equation}
Where 
\begin{multline}
\label{wlt}
\mathcal{L}_{1} = M\Bigl[\frac{i}{2}\left(\psi^{*}\partial_{0}\psi-\psi\partial_{0}\psi^{*}\right)+
-B^{a0}_{0}mx_{a}\psi^{*}\psi\\-\frac{1}{2m}\Sigma^{k}_{a}\Sigma^{l}_{a}\left(\partial_{k}\psi^{*}-iB^{b0}_{k}mx_{b}\psi^{*}\right)\left(\partial_{l}\psi+iB^{c0}_{l}mx_{c}\psi\right)\Bigr]
\end{multline}
and
\begin{multline}
\mathcal{L}_{HE}=\epsilon^{kl}\epsilon_{ab}\bigl[\partial_{l}B^{ab}_{k}+\frac{1}{4}B^{a0}_{k}B^{b0}_{l}+\Lambda^{a}_{k}\partial_{0}B^{b0}_{l}-\Lambda^{a}_{k}\partial_{l}B^{b0}_{0}-\frac{1}{2}\Lambda^{c}_{k}B^{c0}_{0}B^{ba}_{l}-\Lambda^{c}_{k}B^{cb}_{0}B^{a0}_{l}\bigr] 
\end{multline}
which is obtained from (\ref{NR}) by putting $ \Sigma_0{}^k$ = 0 in it.The canonical analysis proceeds in the same way as above.

Performing the canonical analysis, we obtain the following primary constraints: 
\begin{align}
\Omega_{1}=\pi-\frac{Mi}{2}\psi^{*}\thickapprox 0\hspace{.2cm};\hspace{.2cm}
\Omega_{2}&=\pi^{*}+\frac{Mi}{2}\psi \thickapprox 0\notag\\
\Omega^{0}_{a0}=\pi^{0}_{a0}\thickapprox 0\hspace{.2cm};\hspace{.2cm}
\Omega^{a}_{k}&=\pi^{a}_{k}\thickapprox 0\notag\\
\Omega^{\mu}_{ab}=\pi^{\mu}_{ab}\thickapprox 0\hspace{.2cm};\hspace{.2cm}
\Omega^{l}_{b0}&=\pi^{l}_{b0}-\epsilon^{kl}\Lambda^{a}_{k}\epsilon_{ab} \thickapprox 0
\end{align}
The stationary of the primary constraints  $\Omega^{0}_{d0}$, $\Omega^{d0}_{b}$, $\Omega^{j}_{ef}$, $\Omega^{ej}_{f}$  give the following secondary constraints:
\begin{align}
\Phi&= \psi^{*}\psi\thickapprox 0 \notag \\
\Phi_{d}&=\epsilon^{kl}\epsilon_{ad}\partial_{l}\left(\Lambda^{a}_{k}\right)+\frac{1}{2}\epsilon^{kl}\epsilon_{ab}\Lambda^{d}_{k}B^{ab}_{l} \thickapprox 0\notag \\
\Gamma_{ef}&=\epsilon^{kl}\epsilon_{ab}\Lambda^{c}_{k}B^{b0}_{l}\left(\delta^{c}_{e}\delta^{a}_{f}-\delta^{a}_{e}\delta^{c}_{f}\right)  \thickapprox 0\notag \\
\Gamma^{j}_{ef}&=-\frac{1}{2}\epsilon^{kl}\epsilon_{ab}\Lambda^{c}_{k}B^{c0}_{0}\delta^{j}_{l}\left(\delta^{a}_{e}\delta^{b}_{f}-\delta^{b}_{e}\delta^{a}_{f}\right)  \thickapprox0\notag \\
\end{align}
The iteration terminates with the closure of the constraint algebra.

The non-vanishing poisson brackets between the constraints are given by
\begin{align}
\label{pbsplit}
{\{{\Omega}_{1}(x),\Omega_2(y)\}}&=-iM\delta^{2}(x-y)\\
{\{\Omega_{1}(y)},\Omega^{a}_{k}(x)\}&=\frac{i\psi^{*}}{2}M\Lambda^{a}_{k}\delta^{2}(x-y)\\
{\{\Omega_{1}(x),\Phi(y)}\}&=-\psi^{*}\delta^{2}(x-y)\\ 
{\{\Omega_{2}(y)},\Omega^{a}_{k}(x)\}&=-\frac{i\psi}{2}M\Lambda^{a}_{k}\delta^{2}(x-y)\\
{\{\Omega_{2}(x),\Phi(y)}\}&=-\psi\delta^{2}(x-y)\\ 
{\{\Omega^{a}_{k}(x)},\Omega^{l}_{b0}(y)\}&=-\epsilon^{pl}\epsilon_{db}\Lambda^{d}_{k}\Lambda^{a}_{p}\delta^2(x-y)\\
 {\{\Omega^{a}_{k}(x)},\Gamma_{ef}(y)\}&=\epsilon^{jl}\epsilon_{db}\Lambda^{c}_{k}\Lambda^{a}_{j}B^{b0}_{l}\left( \delta^{c}_{e}\delta^{d}_{f}-\delta^{d}_{e}\delta^{c}_{f}\right) \delta^2(x-y)\\
 {\{\Omega^{a}_{k}(x)},\Gamma^{j}_{ef}(y)\}&=-\frac{1}{2}\epsilon^{pj}\epsilon_{db}\Lambda^{c}_{k}\Lambda^{a}_{p}B^{c0}_{0}\left(\delta^{d}_{e}\delta^{b}_{f}-\delta^{b}_{e}\delta^{d}_{f}\right)\delta^2(x-y)\\
{\{\Omega^{0}_{a0}(x)},\Gamma^{j}_{ef}(y)\}&=\frac{1}{2}\epsilon^{pj}\epsilon_{db}\Lambda^{a}_{p}\left(\delta^{d}_{e}\delta^{b}_{f}-\delta^{b}_{e}\delta^{d}_{f}\right)\delta^2(x-y)\\
{\{\Omega^{k}_{ab}(x)},\Phi_{d}(y)\}&=-\frac{1}{2}\epsilon^{pk}\epsilon_{cf}\Lambda^{d}_{p}\left(\delta^{c}_{a}\delta^{f}_{b}-\delta^{f}_{a}\delta^{c}_{b}\right)\delta^2(x-y)\\
{\{\Omega^{l}_{b0}(x)},\Gamma_{ef}(y)\}&=-\epsilon^{pl}\epsilon_{ab}\Lambda^{c}_{k}\left(\delta^{c}_{e}\delta^{a}_{f}-\delta^{a}_{e}\delta^{c}_{f}\right)\delta^2(x-y)\\
{\{\Omega^{l}_{b0}(x)},\Gamma_{j}(y)\}&=-\epsilon^{kp}\epsilon_{ad}\left[ \Lambda^{a}_{j}\delta^{l}_{k}\delta^{d}_{b}\partial^{y}_{l}(\delta^2(x-y))-\frac{1}{4}\Lambda^{b}_{j}\Lambda^{l}_{k}B^{ad}_{p}\delta^2(x-y)-\frac{1}{2}\Lambda^{c}_{j}\delta^{l}_{p}\delta^{d}_{b}B^{ca}_{k}\delta^2(x-y)\right]\\
\end{align}
The complete classification of constraints is summarized in Table 2 below.
\begin{table}[ht]
\caption{Classification of Constraints when $\Sigma_0{}^k = 0$}
\label{tab:Classification of Constraints of the truncated model}
\centering
\begin{tabular}{c c c}
\hline\hline\\

\ & First Class & Second Class \\
 \hline\\
 Primary &  ${\Omega^{0}_{ab}}$ & ${\Omega_{1}}$, ${\Omega_{2}}$, $\Omega^{a}_{k}$, $\Omega^{0}_{a0}$, $\Omega^{k}_{ab}$, $\Omega^{l}_{b0}$, \\[1em]
\hline\\
 Secondary &  & $\Phi$,$\Phi_{d}$, $\Gamma_{ef}$, $\Gamma^{j}_{ef}$ \\[1em]
 \hline\hline
\end{tabular}
\end{table}
The number of fields is 15, the number of first class constraints is 1 whereas there are 20 secondary constraints. So the number of degrees of freedom in the phase space is 8. This is larger  as the physical degrees of freedom. So we see that the model with $\Sigma_0{}^k = 0$ is unable to give the hamiltonian analysis consistently.

Again, we see from table -2 that the number of primary first class constraints is one. So the model predicts one local symmetry as opposed to three physical symmetries!. So taking $\Sigma_0{}^k =0$ also gives incorrect symmetries. Further
investigation shows that the boost symmetries are lost. This connection with boost is indeed remarkable, not only for GGT but also in general.

 \section{ Concluding remarks}
 
 The
 complete Schroedinger field theory interacting with non -relativistic Newtonian gravity has been given in the vierbein
 approach \cite{BMM1}, \cite{ BM1}. This is apparently for the first time a complete Hamiltonian treatment of the problem within the framework of non - relativistic diffeomorphism invariance is reported . A thorough canonical analysis has been given. This analysis was found to be the originator of the full fledged consistency The canonical analysis
 was repeated with time space component of vierbein set to zero. This time anomalous results are obtained in the degrees
 count. The three dimensional gravity assumed here is known to contribute no propagating degree of freedom
 in the relativistic theory. We have shown that this happens for the non relativistic theory as well when GGT directs the
 dynamics. 
 The Hamiltoni ananalysis of a Schroedinger field with  background dynamical Newtonian action is presented in this paper.
  The attempt was entirely successful and is remarkable due to a number of reasons. 
 
\begin{enumerate}

\item

The Schroedinger field theory interacting with gravity has been discussed over a long period of time \cite{SW} due to its applications in the fractional quantum Hall effect. But so far we know ,very few Hamiltonian analysis are available \cite{kluson}. In these works the gravity is responsible only to a coupling with the curved space. It does not have
any dynamics. So our paper is possibly the first paper where the mutual interaction is observable. We have used the
correspondence derived recently between the Riemann Cartan and Newton Cartan spacetime \cite{BM5} in the 
 Einstein - Hilbert action. That this action represents  the Newtonian gravity theory is obvious
\cite{MTW}. We have explicitly checked that the action is invariant under the extended Galilean transformations
here. 

\item

The coupled theory in the flat space is invariant under the extended Galilean Galilean  group of transformations.
No non canonical term is obtained as in \cite{SW} or no additional gauge field in the Newton - Cartan geometry  
  as required. The whole theory follows from a monolithic logical system i.e, the Galilean gauge theory.

 \end{enumerate}

%


\end{document}